# CERTIFIED BINARY SEARCH TREE ON W-TYPES

# ÁRBOL BINARIO DE BÚSQUEDA CERTIFICADO SOBRE W-TIPOS

**PhD. Gustavo Arengas**

**Universidad Nacional de Colombia**
Departamento de Matemáticas
Carrera 30 No. 45-03. Bogotá, Colombia.
Teléfono: 3165000 ext. 13153
E-mail: gearengasr@unal.edu.co

**Abstract:** Software bugs have caused enormous economic and human loss in recent years. Certified programming seeks to solve this problem by developing languages where we can make demonstrations that guarantee that our programs work properly. However, the rapid evolution of modern algorithms constantly forces us to develop new tools for this task. In this article we explore the possibility of defining new types using well-founded types. In particular, we developed a new elimination rule on the terms of the type. We apply these ideas to binary search trees to illustrate the power of the method.

**Resumen:** Los errores de software han causado enormes pérdidas económicas y de vidas humanas en los últimos años. La programación certificada busca solucionar este problema mediante el desarrollo de lenguajes donde podemos hacer demostraciones que garanticen que nuestros programas funcionen adecuadamente. Sin embargo, la rápida evolución de los algoritmos modernos constantemente nos fuerza a desarrollar nuevas herramientas para esta tarea. En este artículo exploramos la posibilidad de definir nuevos tipos mediante tipos bien fundados. En particular desarrollamos una nueva forma de hacer inducción sobre los términos del tipo. Aplicamos estas ideas a los algoritmos basados en árboles binarios de búsqueda para ilustrar la potencia del método.

**Palabras clave:** W-tipo, programación certificada, regla de eliminación, árboles binarios de búsqueda.

## 1. INTRODUCTION

Our lives depend more and more on computers, so when they fail the consequences are disastrous. For example, the cost of software bugs has grown exponentially in recent years: it is estimated that in 2016 it amounted to US$1.1 trillion worldwide (McPeak, 2017, August 2) and that in 2017 that number jumped to US$1.7 trillion (Platz, 2018). Among the most notable cases is that of Knight Capital Group, which lost US$440 million in 45 minutes in 2012 due to a software error that predicted the behavior of the shares (Popper, 2012, August 2); that of Provident Financial that lost £1.7 billion (US$2.4 billion) of its market value in just one day in 2017 due to an error in the software that scheduled its payments (Davies, 2016, August 16); and the U.K.'s Child Support Agency, which due to a bug in the software that handled the payment of subsidies, has cost UK taxpayers more than US$1 billion to date.

Unfortunately, losses are not just financial. Between 1985 and 1987 a fail in the code of the Therac-25, a computer-controlled radiation therapy machine, overdosed and kill five people (Levenson, 1995); in 1991 an error in the internal clock of the Patriot defense system allowed an Iraqi missile to kill 28 soldiers (Skeel, 1992); and in 2009 four passengers were died when their car crashed for a software bug that caused a lag in the anti-lock-brake system (Bensinger, 2009, october 25).

This explains the growing importance that certified programming is taking in recent years. A certification of a program is a formal guarantee that it has some specific properties. For example, it could be demonstrated that the program always terminate, or that it has no safety concerns, or that it meets certain efficiency standards. The basis of this technique is the Curry-Howard correspondence, which states that computer programs and intuitionistic mathematical proofs are the same kind of mathematical object, which is traditionally called type. Thus it is possible to develop programming languages, called proof assistant, that allow us to run programs and check their properties at the same time. Among the best known proof assistants are Agda (Norell, 2009), Coq (Coquand and Huet, 1989), Isabelle (Nipkow and Klein, 2014).

In the proof assistants we generally introduce the new types through inductive definitions (Chlipala, 2013). In this work we want to explore the alternative of introducing them through W-types, which allow us to have more control over the constructors involved. As an application of the exposed ideas, we show how to program the binary search tree algorithm in dependent type theory and we use the W-types tools to demonstrate that this program behaves in the expected way. We will work in a semi-formalized language to facilitate the reading of the article, although the ideas presented can be formalized in proof assistants such as Agda or Coq directly.

## 2. W-TYPES

Throughout this work we will denote with $\mathcal{U}$ the universe of types. Given a type $A$ and a dependent type $B$ on $A$, a family of types indexing by the terms of $A$ (Aspinall and Hofmann, 2005), the type $W_{x:A}B(x)$ denote the well-founded tree with nodes labelled by $A$ and links given by the image of $B$. In type theory, each former is introduced by four rules, which in the case of W-types have the form:

**Definition 1.** (Martin-Löf, 1984)
● $W_{For}$ If $A : \mathcal{U}$ and $B : A \to \mathcal{U}$, then $W_{x:A}B(x) : \mathcal{U}$.
● $W_{Int}$ If $a:A$ and $f : B(a) \to W_{x:A}B(x)$, then $sup(a,f) : W_{x:A}B(x)$.

● $W_{Eli}$ Given $C:W_{x:A}B(x) \to \mathcal{U}$, if for all $a:A$ and $f:B(a) \to W_{x:A}B(x)$, we have that $v:\prod_{y:B(a)}C(f(y))$ implies that $\alpha(a,f,v):C(sup(a,f))$, then for any $w:W_{x:A}B(x)$, exists $ind_W(w, \alpha):C(w)$.
● $W_{Com}$ For any $a:A$ and $f:B(a) \to W_{x:A}B(x)$, we have
$$ind_W(sup(a,f),\alpha) :\equiv \alpha(a,f,\lambda y.ind_W(f(y),\alpha)).$$

$W_{For}$ is a formation rule that establishes when the W former can be used; $W_{Int}$ is an introduction rule that establishes how the new type be inhabited; $W_{El}$ is an elimination rule that state the induction principle to use the type; and $W_{Com}$ is a computation rule that ensures that constructions made by elimination behave as expected.

Among the advantages of introducing definitions through W-types versus working with inductive types are:
● We can enter all inductive types with the only *sup* constructor, instead of entering them one by one.
● We can establish general results that are valid for all W-types while we cannot do the same with inductive types, since there are no precise rules that govern them (Voevodsky *et al.*, 2013).
● Many proof assistants more easily accept definitions introduced via W-types. For example, Coq has trouble accepting definitions of nested inductive types, but these can be entered via W-types (Abbott and Altenkirch, 2004).

It is convenient to think of the terms of a W-type as ordered pairs, the first component of which indicating the arity of the constructor and the second the function that indicates the elements taken by this constructor. In the following proposition we show how to define projections to extract this information.

**Proposition 2.** Let $W := W_{x:A}B(x)$. There are functions:
$$pr_1: W \to A$$
$$pr_2: \prod_{w:W} (B(pr_1(w)) \to W)$$
such that for all terms of the form $sup(a,f)$, $pr_1(sup(a,f)) \equiv a$, $pr_2(sup(a,f)) \equiv f$.

*Proof.* We can define $pr_1$ by taking $C(w) \equiv A$ and $\alpha(x,u,v) \equiv x$ in the $W_{Eli}$ rule. The resulting function has the desired properties thanks to $W_{Com}$. Analogously, we construct $pr_2$ from $C(w) \equiv B(pr_1(w)) \to W$ and $\alpha(x, u, v) \equiv u$.

□

It can be shown that a w-type W is different from empty if and only if it has terms of arity 0 (Martin-Löf, 1984; Voevodsky *et al.*, 2013). In the following we will call these terms as minimals.

**Definition 3.** The minimals terms of $W_{x:A}B(x)$ is the dependent type $isMin: W_{x:A}B(x) \to \mathcal{U}$, given by $isMin(w) :\equiv \neg B(pr_1(w))$.

Definition 1 does not establish any restriction on the arity of the terms that make up a w-type. However, in what follows it is convenient to assume that they all have finite arity.

**Definition 4.** We say that a type **W** is of finite arity if for each $w:\mathbf{W}$, there exists a natural number $n$, such that the number of different terms in $B(w)$ is equal to $n$. In the language of theory, this can be expressed as a dependent type, which we will call *isFiAr*, defined for a W-types $\mathbf{W} \equiv W_{x:A}B(x)$ as

$isFiAr(W) \equiv \prod_{w:W} \sum_{n:\mathbb{N}} \|Br(pr_1(w))\| = n$

where $\|B\|$ is the number of different terms in $B$.

In the finite-arity W-types we can define the complexity of each term as the number of times the constructors have to apply on the minimales to build that term.

**Proposition 5.** Let $W_{x:A}B(x)$ be an finite-arity W-type. Then exists a function
$$comp : W_{x:A}B(x) \to \mathbb{N}$$
definite by:
1. if **isMin**($w$) then $comp(w) :\equiv 0$;
2. if $\neg$**isMin**($w$) then
$comp(w) :\equiv max\{comp(pr_2(w)(b))|b:B(pr_1(w))\}+1$

*Proof.* For $\mathbf{W_{Eli}}$, it is enough to consider a term of the form $sup(a,f)$ and assume that we have already defined $comp(f(y))$ for each $y:B(a)$. The only problem we could have is that $max\{comp\ f(y)|y:B(a)\}+1$ was not a number. But since $W_{x:A}B(x)$ is of finite arity, we are calculating the maximum of a finite amount of natural numbers, which always exists.

□

$\mathbf{W_{Eli}}$ is the basic rule for defining functions on a W-type. However, sometimes it is easier to make an induction on the complexity of the terms. The following proposition gives us this possibility.

**Proposition 6.** Let $W_{x:A}B(x)$ be a W-type and $P:A \to \mathcal{U}$ be a dependent type such that exists:
1. $\alpha: \prod_{w:C(0)} P(w)$
2. $\beta: \prod_{n:\mathbb{N}} \left(\prod_{w:C(n)} P(w) \to \prod_{w:C(n+1)} P(w)\right)$

Then exists $\gamma: \prod_{w:\mathbf{W}} P(w)$.

*Proof.* Let $Q : \mathbb{N} \to \mathcal{U}$ be defined by
$$Q(n) :\equiv \prod_{w:C(n)} P(w).$$
The hypotheses 1 and 2 mean, respectively, that $Q$ satisfies case 0 and case successor of an induction proof. Thus, by the rule of elimination of $\mathbb{N}$ there is a function
$$ind_{\mathbb{N}}(Q, \alpha, \beta, \_) : \prod_{n:\mathbb{N}} Q(n) \equiv \prod_{n:\mathbb{N}} \prod_{w:C(n)} P(w)$$
Thus, we can define
$$\gamma(w) :\equiv ind_{\mathbb{N}}(Q, \alpha, \beta, comp(w))(w).$$
□

## 3. BINARY SEARCH TREE

Modern computers make an immense amount of information available to users. Search algorithms allow us to find the information we want efficiently. Among these, those who use binary search trees (Knuth, 1998) have played a leading role for decades, to the point of being considered among the most fundamental in computer science.

**Definition 7.** (Sedgewick and Wayne, 2011) A binary search tree is a binary tree where each node has a key belonging a totally ordered type K and satisfies that the key in any node is larger than the key in all nodes in that node's left subtree and smaller than the keys in all nodes in that node's right subtree.

To program the algorithm that makes use of binary search trees in dependent type theory, we first introduce binary trees.

**Definition 8.** The binary tree type **BT** can be introduced as the W-type with the following constructors:
- $e$ of arity zero, to build the empty tree;
- $cons(k,v,\_,\_)$ of arity two, for each k in K and v in V. Given two trees l, r in **BT**, $cons(k, v, l, r)$ is the tree in BS that has at the root the key k associated with the value k, a l as a left subtree and a r as a right subtree.

More formally, in terms of definition 1, **BT** is the W-type $W_{x:A}B(x)$, where $A \equiv KxV+1$ (the coproduct of KxV and the unit type $\mathbf{1} \equiv \{*\}$), and $B$ is the dependent type definite by $B(*) \equiv 0$ (the empty type), $B(k,v) \equiv 2$ (the type of booleans).

Note that not all binary trees satisfy definition 7. We will later define binary search trees as the BT subtype that have a certain additional property. Meanwhile we define the functions on which the algorithm is based. First of all, we would like to associate values v in V to the keys located in the nodes of the binary trees.. We will do this with the *insert* function.

**Definition 9.** *insert* is a function of type KxVx**BT**→**BT,** which given a key x, value b and a tree t, construct a tree where x is associated with b. Again by Curry-Howard, we can fix x and b, and focus on using W-elimination to build functions of type **BT**→**BT**:

(I.1) if t ≡ *e*, then
$$insert(x,b,e) \equiv cons(x, b, e, e).$$
(I.2) if t ≡*cons*(k, v, l, r), we work by cases about the possible order relation of x and k:

(I.2.1) if x<k, then *insert*(x,b,t) ≡ *cons*(k, v, *insert*(x,b,l), r);

(I.2.2) else, if x>k, then *insert*(x,b,t) ≡ *cons*(k, v,l, *insert*(x,b,r);

(I.2.3) else (x≡k), then *insert*(x,b,t) ≡ *cons*(k, b,l,r).

We also need to define a *search* function that gives us the value in V associated with a given k in K, if that value exists. It is convenient to assume that type V has a certain term, called *default*, which will give us the *search* function when it fails.

**Definition 10.** *search* is a function of Kx**BT** to V, which yields the value associated with the key x in the tree t. By Curry-Howard, a function in AxB→C is the same that a function in A→B→C, we can use W-elimination (**W**_{Eli}) to build families of functions in **BT**→V for each fixed x value. We work in each of the constructor that define **BT**:

(S.1) if t ≡ *e*, the empty tree, then *search*(x,t) ≡ *default*.

(S.2) if t ≡*cons*(k, v, l, r), we work by cases about the possible order relation of x and k:

(S.2.1) if x<k, then *search*(x,t) ≡ *search*(x,l);

(S.2.2) else, if x>k, then *search*(x,t) ≡ *Search*(x,r);

(S.2.3) else (x≡k), then *search*(x,t) ≡v.

*insert* and *search* work almost like inverse functions. For example, we can demonstrate that if we associate the value *v* to the key *k* in the tree *t* through *insert*, and then request the value associated to *k* in the new tree by *search*, we obtain *v*. In terms of type theory, we can express this fact with the following type.

**Proposition 11.**
$$\prod_{x:K} \prod_{b:V} \prod_{t:BT} search(x, insert(x,b,t)) \equiv b$$

*Proof.* By induction on t complexity (Proposition 6):

• if *comp*(t)≡0 then *t*≡*e* and thus:
*search*(*x*,*insert*(*x*,*b*,*e*))≡*search*(x,*cons*(*x*,*b*,*e*,*e*))
≡ b
where the first equation is deduced by (I.1) and the second by (S.2.3).

• if *comp(t)*≡n+1 then *t*≡*cons(k, v, l, r)*, and we work by cases on the order relation of x,k:

-*(x<k)* In this case we have:
*search*(x,*insert*(x,b,*cons*(k,v,l,r)))
≡*search*(x,*cons*(k,v,*insert*(x,b,l),r)   by (I.2.1)
≡*search*(x,*insert*(x,b,l))   by (S.2.1)
≡b,   by induction hypothesis.

-*(x>k)* Analogous to the previous case.

−*(x≡k)* Finally, here we have the equations:
*search*(x,*insert*(x,b,*cons*(k,v,l,r)))
≡*search*(x,*cons*(k,v,l,r))   by (I.2.3)
≡b,   by (S.2.3).

□

To build the binary search trees, we will assume that in the order of K we can define the predecessor and successor functions.

**Definition 12.** *K* is an order with predecessor and successor if for all *k:K*, with *k* different from *default*, there are *P(k)* and *S(k)* with the following properties:
• *P(k)<k* and there is no *z* such that *P(k)<z<k*.
• *k<S(k)* and there is no *z* such that *k<z<S(k)*.

The property we define below, *rank(n,t,m)*, will be true when the keys of the left subtree of *t* are greater than *n* and less than the key of the root of *t*, and at the same time the keys of the right subtree of *t* are greater than that of the root and less than *m*. The fundamental fact is that the *rank* function recursively goes through all the nodes of *t*, thus catching the idea of definition 7.

**Definition 13**. *rank*:Kx**BT**$_X$K→**2** is an function recursively constructed by:
(R.1) *rank*(n,e,m):≡⊤ iff n≤m.
(R.2) if t≡*cons*(k,v,l,r) then
*rank*(n,t,m)≡*rank*(n,l,P(k))∧*rank*(S(k),r,m)
.

The following lemmas will be useful later.

**Lemma 14.** If *rank(n,t,m)*≡⊤ and `u≤n, m≤v` then *rank(u,t,v)*≡⊤.

*Proof.* By induction on the complexity of the tree *t*:
- if `comp(t)≡0` then `t≡e.` We have `n≤m` because *rank(n,e,m)*≡⊤. So *rank(u,e,v)*≡⊤ by (R.1) and `u≤n≤m≤v.`
- if `comp(t)≡n+1` then *t*≡*cons(k,v,l,r).* We have (R.2) that:
*rank(n,t,m)*≡*rank(n,l,P(k))*∧*rank(S(k),r,m).*
As *rank(n,l,P(k))*≡⊤, `then by induction hypothesis` *rank*(u,l,P(k))≡⊤. `Analogously,` *rank(S(k),r,m)*≡⊤. `Then, by the definition of the connective,` *rank(u,cons(k,v,l,r),v)*≡⊤.

□

**Lemma 15**. If *rank(n,t,m)* and `n≤x≤m`, then *rank(P(n),insert(x,b,t),S(m)).*

*Proof.* By induction on the complexity of the tree *t*:
- if `comp(t)≡0`, then `t≡e` and by (I.1, def. 10) *insert(x,b,t)*≡*cons(x,b,e,e).* By (R.1, def. 13) *rank*(n,e,m) ≡ *rank*(s(n),e,m) ≡ ⊤ and thus *rank*(n,*insert*(x,b,t),m)≡⊤.
- `if comp(t)≡n+1, then t≡`*cons*(k,v,l,r) and we work by cases about the possible order relation of x and k:
(x<k) then *insert*(x,b,t) ≡ *cons*(k, v, *insert*(x,b,l), r). *rank*(s(n),r,m)≡⊤ because *rank(n,t,m)*≡⊤ and
*rank*(n,*insert*(x,b,l),*m*)≡⊤ `by induction hypothesis.                    Thus` *rank*(n,*insert*(x,b,t),m)≡⊤ `by definition (R.2, definition ).`
(x>k) Analogous to the previous case.
(`x≡k`) then *insert*(x,b,t) ≡ *cons*(k,b,l,r) and thus
*rank*(n,*insert*(x,b,t),*m*)≡*rank*(n,l,m)∧*rank*(s(n),r,m)≡*rank*(n,t,m)≡⊤.

□

Binary search trees will be those binary trees that behave well with the rank operation.

**Definition 16.** The binary search tree type **BST** is the subtype of **BT** well behaved with the range function:
  **BST**≡$\sum_{t:BT} \sum_{n:K} \sum_{m:K}$ *rank*(n,t,m).

The following propositions show that, as expected (Knuth, 1998; Sedgewick and Wayne, 2011), binary search trees can be built from the empty tree by inserting pairs of keys and values.

**Proposition 17.** The empty tree *e* has type **BST**.

*Proof.* It is enough to take a pair *n,m*:K `with n≤m, because then` *rank(n,e,m)*≡⊤ `for (R.1) in definition 13.`

□

**Proposition 18.** For all x:K, b:V, t:**BT** if t:**BST** then *insert(x,b,t)*:**BST.**

*Proof.* If t:**BST** then exists n',m':K such that *rank*`(n',t,m') is inhabited. Let's define n = n'`∧`x and m = m'`∨`x. As` *rank(n',t,m')* `and n≤n', m'≤m then` *rank(n, t, m)*, by lemma 14. As `n≤x≤m`, then *rank*(P(*n*),*insert*(x,b,t),S(*m*)) by lemma 15.

□

## 4. CONCLUSIONS

Although many of the concepts of certified programming are still under development, it is expected to play a leading role in industry and finance in the near future. In this article we wanted to explore the possibility of working with W-types instead of inductive types, which are more commonly used. In particular, a new way of doing induction on W-types, the complexity induction of the term, was developed and we illustrate its usefulness through various propositions.

We decided to illustrate the power and flexibility of the method by showing how the binary search tree algorithm can be introduced into dependent type theory. We first define the type of binary trees, along with the pseudocodes of the *insert* and *search* functions. We show as an example how that if we insert a value to a key in a given tree and then *search* that key, we get the inserted value. Later we introduce the binary search trees as the subtype of the binary trees that satisfy the classical property: the key associated with each node is greater than those located in the left subtree and less than those located in the right subtree. We demonstrate that all trees built of the empty tree by repeated application of the insert function belong to this subtype.

Many questions remain in the field of w-types, in particular, and in that of certified programming, in general. Some research suggests the possibility that many types, such as identity types and through these higher inductive types (Lumsdaine and Shulman, 2019), can be defined as w-types or adequate generalizations of these, greatly extending the applications of this type class. In the field of certified programming, new testing techniques are being developed, such as random testing (Lampropoulos, 2018), which opens up interesting connections to machine learning. We hope to explore some of these paths in future work.